\documentclass[twocolumn]{aastex62}

\received{}
\revised{}
\accepted{}
\submitjournal{ApJ}

\shorttitle{}
\shortauthors{}

\begin{document}

\title{The Period Change and Long-Term Variability of Cyg~X-3}

\correspondingauthor{Igor I. Antokhin}
\email{igor@sai.msu.ru}

\author[0000-0002-3561-8148]{Igor I. Antokhin}
\affil{Moscow Lomonosov State University, \\
Sternberg State Astronomical Institute, \\
119992 Universitetsky prospect, 13, Moscow, Russian Federation}

\author[0000-0001-5595-2285]{Anatol M.Cherepashchuk}
\affil{Moscow Lomonosov State University, \\
Sternberg State Astronomical Institute, \\
119992 Universitetsky prospect, 13, Moscow, Russian Federation}

\begin{abstract}

By using available archival X-ray data, we significantly extended the list of times of X-ray minima. The new list includes 65 data points obtained by critically re-analyzing {\em RXTE} ASM data, 88 data points based on observations by {\em MAXI}, and two data points based on  observations by {\em SUZAKU} and {\em AstroSat}. Analyzing the data along with times of X-ray minima available from the literature, we provide the most accurate estimate of the rate of period change to date. We do not confirm existence of a second derivative of the orbital period suggested by some authors earlier. Instead, we find that the changes in the period can be fit by a sum of quadratic and sinusoidal functions. The period of sinusoidal variations is $15.8$~yr. They can be related either to apsidal motion in the close binary with eccentricity $e\simeq 0.03$ or to a presence of a third body with the mass of about $0.7$~M$_\odot$ located at a distance $\sim 16$~au from the close binary. We also detect irregular and abrupt changes in the residuals between the best fit ephemeris and the data.  While we discuss possible reasons for the changes, their origin remains unclear. A tentative period of about 188 days in X-ray flux variations was found. Such a period could be attributed to a small precessing disk around the compact object.

\end{abstract}


\section{Introduction} \label{sec:intro}

Cyg~X-3 is a rare X-ray binary system consisting of a WN4-8 star and a compact object \citep{vankerk92, vankerk96}. The only two other (much less studied) binaries of this type are IC10~X-1 (WNE+C, $p=1^d.5$, \citealp{prest07}, \citealp{silver08}) and NGC300~X-1 (WN5+C, $p=1^d.3$, \citealp{crow10}). The orbital period of Cyg~X-3 is about 4.8~hr. Little is known about the system inclination and mass functions. \cite{hanson00} discovered an absorption detail in the IR spectrum, which radial velocity {\em may} reflect orbital motion of the WR component. \cite{vilhu09} found Doppler shifts of some X-ray emission lines which {\em may} reflect orbital motion of the compact object. Critically analyzing these data, \cite{zd13} estimated the mass ratio in the system \mbox{$M_{\rm C}/M_{\rm WR}=0.23^{+0.09}_{-0.06}$}. Then, using a relation between the binary period derivative and the mass-loss rate of the WR component, and relationship between the mass-loss rate and the mass for WR stars of the WN type, they were able to estimate the mass of the WR component, and, subsequently, the mass of the compact component. This latter turned out to be $M_{\rm C}=2.4^{+2.1}_{-1.1}{\rm M_\odot}$. Recently, \cite{kol17} showed that the IR lines found by \cite{hanson00} most likely do not reflect the motion of the WR component and are instead originated in the WR wind. Thus, the conclusion of \cite{zd13} is refuted and the origin of the compact object remains unclear -- it can be either a neutron star (NS) or a black hole (BH).

To shed some light on properties of the components in \mbox{Cyg~X-3}, we undertook extensive JHK observations of the binary with the 2.5~m telescope of the MSU Caucasian observatory. The results of this study will be published in a subsequent paper. In the current paper we focus on refining the rate of period change and studying the long term variability (if any) of \mbox{Cyg~X-3}.

The change of the binary period has been studied by many authors since 1975 (see \cite{bh17} and references therein; the full list of previous papers on the subject is also provided in the references to Table~\ref{table1}). These authors used data obtained with various X-ray satellites. The data for most satellites cover only limited number of orbital cycles. Added to that is strong irregular variability of the source on top of its orbital variations. {\em RXTE} ASM and {\em MAXI} missions, on the other hand, have a big advantage of providing continuous monitoring of \mbox{Cyg~X-3} on time span of 18 years. We take this opportunity to refine our knowledge about the period evolution. In Section~\ref{sec:data} we describe the archival data analyzed in this paper. In Section~\ref{sec:method}, the method for finding the local epoches of X-ray minima and their errors is explained. Section~\ref{sec:pdot} presents the results of fitting these data by various models of period change. In Section~\ref{sec:sine} we discuss a possible nature of the discovered sinusoidal variations of the period. Irregular and abrupt changes in the (O-C) residuals are briefly discussed in Section~\ref{sec:irreg}. In Section~\ref{sec:longterm}, we present and discuss the results of a search for long term variability of X-ray flux. Section~\ref{sec:concl} contains a summary of our results.

\section{Archival Data}\label{sec:data}

\subsection{RXTE {\em ASM}}

The ASM on board the {\em RXTE} satellite \citep{levine96} had 3 scanning cameras with large field of view (FOV, $6\arcdeg\times 90\arcdeg$ for each camera). A single exposure was 90~s, then the telescope moved to another position. The light curve in the total energy range 1.3-12~keV was downloaded from {\em RXTE} Guest Observer Facility provided by HEASARC. The total number of data points was 97996, covering the period from January 1996 to December 2011. Note that some of the measured fluxes are negative, probably due to incorrect background subtraction when the source flux was very low. These measurements were excluded from the analysis.

\subsection{SUZAKU}

\mbox{Cyg~X-3} was observed by {\em SUZAKU} in November 2006 for about 50~hr. The data useful for extracting the light curve were obtained with three XIS (X-ray Imaging Spectrometers) instruments (XI0, XI1, XI3). The full FOV is $18\arcmin\times 18\arcmin$. The source, however, was observed in a burst mode with the window size $4.5\arcmin\times 18\arcmin$. The energy range covered by XIS is 0.2-12~keV. We downloaded the event lists and auxiliary files from the HEASARC archive facility. The data were processed with the {\em Ftools} package. To increase signal to noise ratio, we combined the light curves (in the total energy range) from XI0 and XI3 (XI1 cannot be combined with other XIS instruments as its CCD has a distinctly different response). The light curve contains 621 measurements.

\subsection{MAXI {\em SCAN Data}}
	
The {\em MAXI} on board International Space Station (ISS) is a scanning X-ray telescope launched in 2009 \citep{maxi09}. The fields of view of its two instruments, GSC and SSC ($1\arcdeg.5\times 160\arcdeg$ and $1\arcdeg.5\times 90\arcdeg$ respectively) are oriented perpendicularly to ISS orbit and thus allow to observe many objects at once nearly every ISS orbit (about 90~minutes), subject to ISS orbit precession. An object located along a great circle stays in the FOV for 45~s, slanted objects stay longer. The {\em MAXI} web site provides links to light curves averaged over various time intervals, but also a link to the so-called SCAN data which represent individual observations obtained within one ISS orbit. We downloaded GSC SCAN data for \mbox{Cyg~X-3} from the {\em MAXI} web site. The data  cover the period from 2009 August 15 to 2018 March 24, 25042 measurements in total. Among them, about 23000 have exposure times from 47 to 53~s, the rest of measurements have exposures from about 20 to 190~s. For each measurement, the data consist of fluxes in 2-20, 2-4, 4-10, and 10-20~keV bands. To study the period change and long-term variability, we used the data for the total band 2-20~keV.

\subsection{AstroSat}

\mbox{Cyg~X-3} was observed by {\em AstroSat} for 1.5 days in November 2015. The useful data were obtained by the SXT instrument, the energy range is 0.3-8~keV. We downloaded the light curve using a link provided in \cite{bh17}. The number of measurements is 192.

The times of observations for all archival data were corrected to the solar system barycenter.

\subsection{Data segments}

The data from {\em SUZAKU} and {\em AstroSat} cover short time intervals and thus provide only one local epoche of X-ray minimum each. To determine local epoches for {\em RXTE} ASM and {\em MAXI} data, the corresponding data sets were divided by segments as follows:

\begin{enumerate}
 \item The average length of a segment should be relatively small (tens of days).
 \item The object should be either in high or low state within a segment, without transitions between the two.
 \item A segment should not include strong flares significantly affecting the shape of the light curve.
\end{enumerate}

These rules, being applied, resulted in one data segment for {\em SUZAKU} and {\em AstroSat} (each), 65 data segments for {\em RXTE} ASM, and 88 data segments for {\em MAXI}. {\em RXTE} ASM data are rather sparse so the length of a segment varied from about 50 to 90 days. For {\em MAXI}, these were from about 15 to 30 days.

\section{The method to determine a local epoche and its error}\label{sec:method}

\begin{figure}
\centering
\plotone{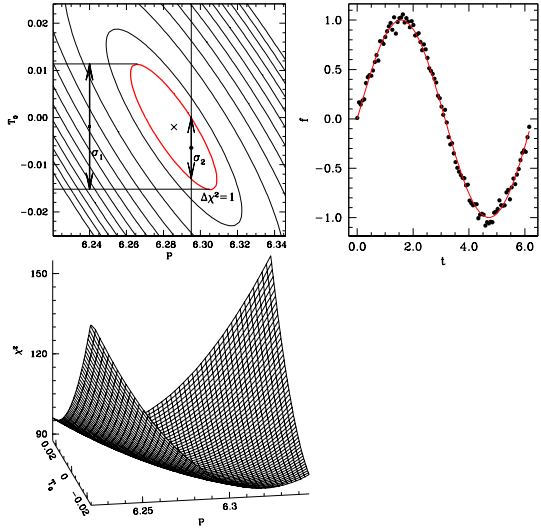}
\caption{Illustration of correlation between local epoche and period. Top right: the simulated data (black dots) and the best-fit sine function (solid red line). Bottom left: $\chi^2$ surface as function of the local period and epoche. Top left: the contour plot of the $\chi^2$ surface. See text for details.}
\label{fig1}
\end{figure}

To determine a local epoche of X-ray minimum for a data segment, early authors fitted the observed light curve by a sine function. However, since the work of \cite{klis89} is it common to fit the observed light curve by the template light curve suggested in this paper. The phase of the template is adjusted so as to coincide to the sinusoidal minimum phase, to avoid systematic shifts. In the most general case, such fitting has four parameters: the scaling factor for the amplitude of the template, its mean flux, the local period and epoche for a data segment. Most previous authors searched for all these parameters.

Note that the authors of the two latest papers on the subject (\citealt{singh02}, \citealt{bh17}) used a fixed value of a local period. \cite{singh02} used a constant period for all data segments they analyzed. \cite{bh17} used two methods -- a constant period and a local period computed from previously found rate of period change. In both cases the value of the local period was fixed while searching for the local epoche. These methods are subject to systematic error in the value of the local epoche and to underestimation of its uncertainty. Indeed, when searching for a period and epoche of a periodic function, the values found by fitting a model are correlated. This is illustrated in Fig.~\ref{fig1}. It shows the results of fitting a sine function computed at 100 equally separated phases of a single cycle, with added Gaussian noise ($\sigma=0.05$), by a sine model. A valley in the $\chi^2$ surface is clearly seen. The position of the best-fit parameters is shown by the cross in the top-left plot. The 1-$\sigma$ error of the epoche $T_0$ is shown by thick arrows in the same plot ($\sigma_1$). As the deviations of the data from the model in this case represent purely Gaussian noise, $\sigma_1$ is given by the extent of the ellipse corresponding to the increase of $\chi^2$ by 1.0 relative to its minimal value (the ellipse is shown by the thick red line). When using a fixed value of the period (e.g. shown by the vertical line in the top left plot in Fig.~\ref{fig1}), the optimal value of $T_0$ (shown by the solid round dot in Fig.~\ref{fig1}) may be systematically shifted from the overall optimal value, and its error $\sigma_2$ is underestimated. 

Intrinsic variations of the light curve in most cases make the value of $\chi^2$ larger than the number of d.o.f.. In our study, a typical reduced $\chi^2_\nu$ was about $3$, as in most of previous studies. Thus, formal error estimates of the local epoche values are unreliable. To avoid this problem, many previous authors artificially increased data errors such that the reduced $\chi^2_\nu$ became equal to unity and then estimated the 1-$\sigma$ uncertainty of the local epoche from the $\Delta\chi^2=1$ principle. However, this method is statistically incorrect. It can be used only if one is certain that deviations of the data from the model have purely Gaussian probability distribution, which is definitely not the case for \mbox{Cyg~X-3}. This is why even after artificially increasing data errors, the error of a local epoche obtained with the $\Delta\chi^2=1$ method, is underestimated. To get a more reasonable value of the error, some authors quadratically added a fixed error \citep[e.g. $0.002$,][]{kitamoto95}, to compensate for systematic component in the deviations of the data from the model. The validity of such approach is doubtful.

Also note that \cite{singh02} and \cite{bh17} used another method to estimate the errors of the local epoches. They searched for the best value of a local epoche by cross-correlating the template light curve with the data and then by fitting the peak of the cross-correlation function by a Gaussian or parabola. The position of the Gaussian/parabola maximum was taken as the local epoche value, and the error of this position (that is the error of the approximation) as the error of the local epoche. This procedure of estimating the errors is generally not valid. The error of the peak position is defined by the quality of the data and has nothing to do with the error of the parametric approximation of the peak. The former can be used solely if one proves that the error of the peak position is much smaller than the error of its parametric approximation. This is what was assumed by \cite{bh17}, without any serious grounds. Moving ahead, we can say that our error estimates for the same data as those used by \cite{bh17}, are about two times larger than theirs.

Concluding, we can say that searching for a local epoche by using a fixed value of the local period may lead to systematic errors in its value. Using the $\Delta\chi^2=1$ method underestimates the error of this value, due to systematic deviations of the data from the template light curve. Using the error of a parametric approximation of the cross-correlation function as the error of the local epoche is not justified. For this reason, in the current work we not only added the {\em MAXI} and {\em SUZAKU} data to the whole data set, but also re-analyzed the {\em RXTE} ASM and {\em AstroSat} data. All four parameters mentioned above, were searched for when fitting observed light curves by the template. The Nelder-Mead method was used as the minimization routine.

\begin{figure*}
\centering
\plotone{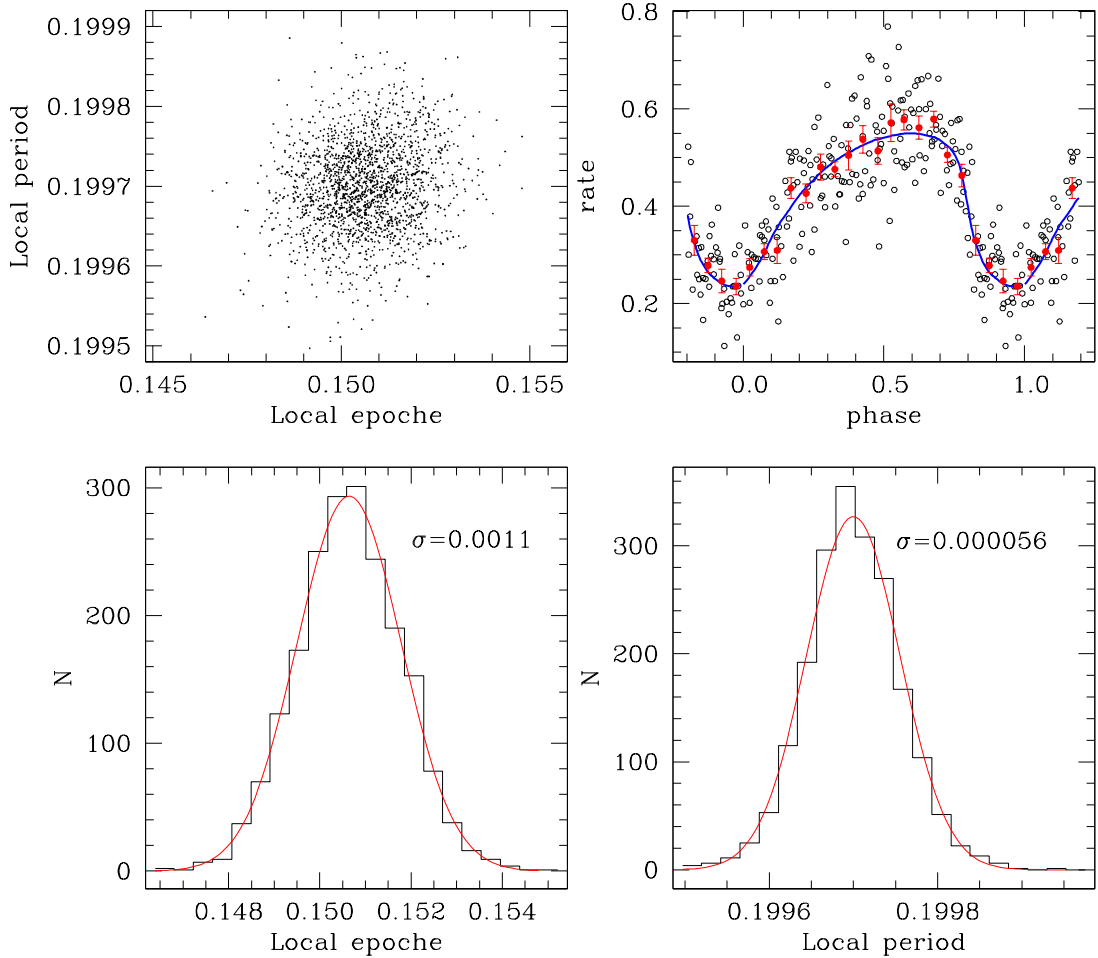}
\caption{An example of the {\em Bootstrapping residuals} method for a {\em MAXI} data segment. Top left: distribution of the values of local epoche and period, each of 2000 dots is the result of a single fit of a synthetic data set. Top right: The best-fit template light curve (solid blue line), the observed data (open dots, the errors of the individual data points are not shown to avoid clutter), the mean observed light curve (red solid dots). Bottom left: the empirical probability function for the local epoche (the histogram). Bottom right: same for the local period. Red lines in bottom plots show Gaussian functions with the mean and $\sigma$ values of the local epoche and period computed from 2000 Monte Carlo simulations.}
\label{fig2}
\end{figure*}

To estimate the error of a local epoche, we used a variant of Monte Carlo simulations known as the {\em Bootstrapping} method \citep{davison97}, in its {\em resampling residuals} version. The method goes as follows. Let us designate the data points to fit as $x_i$, $y_i$, (i=1,...,n), where $x_i$ is the independent variable.

\begin{enumerate}
 
 \item Fit the model, compute the fitted values $y^f_i$, and retain residuals $r_i=y_i-y^f_i$.

 \item Create a new synthetic data set by adding a randomly resampled residual to $y^f_i$: $y^s_i = y^f_i+r_j$, where $j$ is selected randomly from the list ($1, ..., n$) for every $i$.
 
 \item Refit the model using the just created synthetic data set $y^s_i$ and retain the obtained values of the parameters (in our case the local epoche and period).
 
 \item Repeat steps 2 and 3 a large number of times $N$.
 
\end{enumerate}

After the simulations are finished, the obtained $N$ values of every parameter can be used to construct its empirical probability distribution and to estimate various distribution parameters such as standard deviation of a parameter etc. The advantage of this method over $\Delta\chi^2=1$ is that it does not require the deviations of the data from the model to be Gaussian. The only requirement is that the probability distribution of residuals does not vary much in the vicinity of the parameter's true value (i.e., the probability distribution of the residuals with our best fit template is about the same as it would be if the template were computed with the (unknown) true values of the local epoche and period). Of course, such uncertainty estimate is still based on the assumption that the model is accepted.

Our numerical experiments have shown that 2000 simulations were sufficient to get reasonable estimates of $\sigma$ for a local epoche and period. Further increasing the number of simulations did not change the obtained values. In Fig.~\ref{fig2} an example of simulations for one {\em MAXI} data segment is shown. In most cases, the shape of the empirical probability functions is nearly Gaussian, with typical excess kurtosis values from $0$ to $\sim 0.5$. Table~\ref{table1} lists all data on local epoches and orbit numbers (in increasing order) gathered from the literature (115 values) and obtained in the current study (155). The local epoches determined in the current study approximately refer to middle Julian dates of the data segments. The Table does not include the local periods as their errors are usually quite large (see Fig.~\ref{fig2}), making them not too useful.

\begin{deluxetable}{lccc}
 \tablecaption{Local epoches and orbit numbers.\label{table1}}
 \tablehead{
   \colhead{Local epoche $T_n$} &
   \colhead{Error in local epoche} &
   \colhead{Number of} &
   \colhead{Ref.} \\
   \colhead{(MJD)} &
   \colhead{(days)} &
   \colhead{orbital cycle $n$} &
   \colhead{ }
 }

 \startdata
   40949.4201\tablenotemark{a} & 0.0162  &   0 & 1 \\
   40987.1629                  & 0.0037  & 189 & 1 \\
   40987.9555                  & 0.0064  & 193 & 1 \\
   40991.1625                  & 0.0092  & 209 & 1 \\
   41022.4995\tablenotemark{a} & 0.0140  & 366 & 1 \\
 \enddata

 \tablenotetext{a}{The original paper contained a typographic error in this number; corrected by \cite{elsner80}.}
 
 \tablerefs{
   (1) \cite{leach75};
   (2) \cite{mason79};
   (3) \cite{par76};
   (4) \cite{manzo78};
   (5) \cite{lamb79};
   (6) \cite{elsner80};
   (7) \cite{klis81};
   (8) \cite{klis89};
   (9) \cite{kitamoto87};
  (10) \cite{kitamoto92};
  (11) \cite{kitamoto95};
  (12) \cite{singh02};
  (13) Current study.
 }

 \tablecomments{Table~\ref{table1} is published in its entirety in the machine-readable format. A portion is shown here for guidance regarding its form and content.}
 
\end{deluxetable}

\section{The period change}\label{sec:pdot}

\begin{figure}
\centering
\plotone{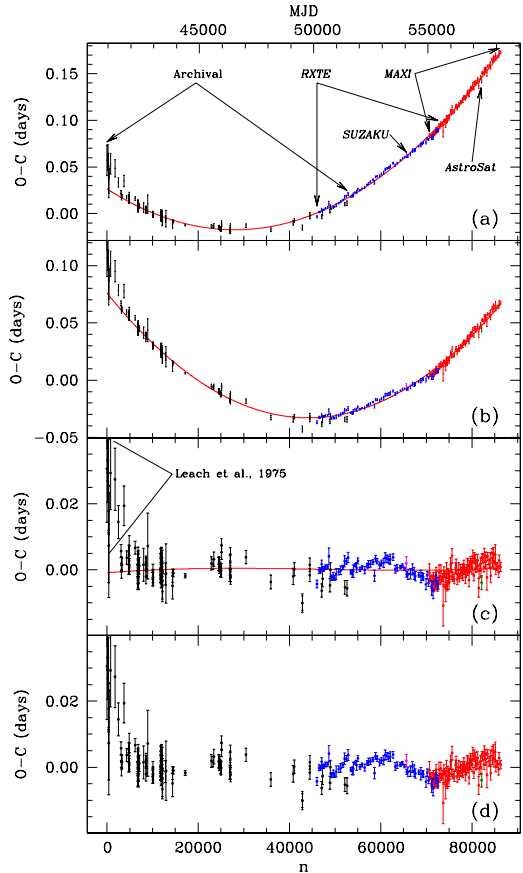}
\caption{Various fits to local epoches of \mbox{Cyg~X-3}. Archival data are shown by black dots, {\em RXTE} ASM data by blue dots, {\em MAXI} data by red dots, {\em SUZAKU} and {\em AstroSat} data points by magenta and green dots respectively. (a) Residuals of the linear model used by \cite{singh02}, shown for reference. The arrows show the epoche intervals covered by archival data and various X-ray missions. The solid red line is the best-fit quadratic model. (b) Residuals of the linear model obtained in the current study. (c) Residuals of the quadratic model obtained in the current study. The solid red line is the best-fit cubic component. (d) Residuals of the cubic model obtained in the current study.}
\label{fig3}
\end{figure}

\begin{table}{l}
 \caption{Ephemeris of different models.\label{table2}}
 \begin{tabular}{l}  
  \hline
  \hline
  Linear ephemeris: $T_n =  T_0 + Pn$ \\
  $\chi^2 = 121270$ for 268 d.o.f \\
  $T_0 = 40949.31465 \pm 0.00021$ MJD\\
  $P = 0.1996895307 \pm 0.0000000025$ d \\
  \hline
  Quadratic ephemeris: $T_n =  T_0 + P_0n + cn^2$ \\
  $\chi^2=946.1$ for 267 d.o.f \\
  $T_0 = 40949.3901 \pm 0.0003$ MJD\\
  $P_0 = 0.199684602 \pm 0.000000015$ d \\
  $c = (5.60 \pm 0.02) \times 10^{-11}$ d \\
  \hline
  Cubic ephemeris: $T_n =  T_0 + P_0n + cn^2 + dn^3$ \\
  $\chi^2=945.9$ for 266 d.o.f \\
  $T_0 = 40949.3903 \pm 0.0005$ MJD\\
  $P_0 = 0.19968458 \pm 0.00000006$ d \\
  $c = (5.67 \pm 0.14) \times 10^{-11}$ d \\
  $d = (-4.6 \pm 9.8) \times 10^{-18}$ d \\
  \hline
  Quadratic ephemeris without \cite{leach75} data: \\
  $\chi^2=782.7$ for 267 d.o.f \\
  $T_0 = 40949.3895 \pm 0.0003$ MJD\\
  $P_0 = 0.199684625 \pm 0.000000015$ d \\
  $c = (5.58 \pm 0.02) \times 10^{-11}$ d \\
  \hline
  Quadratic + sinusoidal ephemeris \\
  without \cite{leach75} data: \\
  $T_n =  T_0 + P_0n + cn^2 + a_s\sin(2\pi/P_s(n-n_{0s}))$ \\
  $\chi^2=495.7$ for 252 d.o.f \\
  $T_0 = 40949.39072 \pm 0.00002$ MJD\\
  $P_0 = 0.19968458145 \pm 0.000000002$ d \\
  $c = (5.616 \pm 0.002) \times 10^{-11}$ d \\
  $\dot{P} = (5.625 \pm 0.002) \times 10^{-10}$ \\
  $a_s = 0.00208 \pm 0.00001$ d \\
  $P_s = 28844 \pm 4$ orbital cycles ($15.77$ yr)\\
  $n_{0s} = 50659 \pm 29$ \\
  \hline
  \hline 
 \end{tabular}
\end{table}

In Fig.~\ref{fig3} the linear, quadratic and cubic model fits to local epoches of \mbox{Cyg~X-3} are shown. The parameters of the fits are listed in Table~\ref{table2}. $\chi^2$ values in all cases are significantly higher than the number of d.o.f., so the errors of the parameters are rather formal. Several previous authors \citep[e.g.,][]{kitamoto95} argued that a cubic model provided a better fit than the quadratic one. Our results clearly show that this is not the case. It is immediately evident from Fig.~\ref{fig3}c that residuals of a quadratic model have sinusoidal shape, except for 12 data points in the beginning of the whole observation interval (see Fig.~\ref{fig3}c). These 12 data points show strong deviations in all models, and this was also the case in all previous studies. The local epoches are from \cite{leach75}, one of the first papers on the subject. The authors obtained local epoches by fitting a sine function to {\em Uhuru} X-ray light curves. As these data points clearly show large systematic deviations from any model, we repeated the quadratic fit excluding these data. The fit parameters are also shown in Table~\ref{table2}.

\begin{figure}
\centering
\plotone{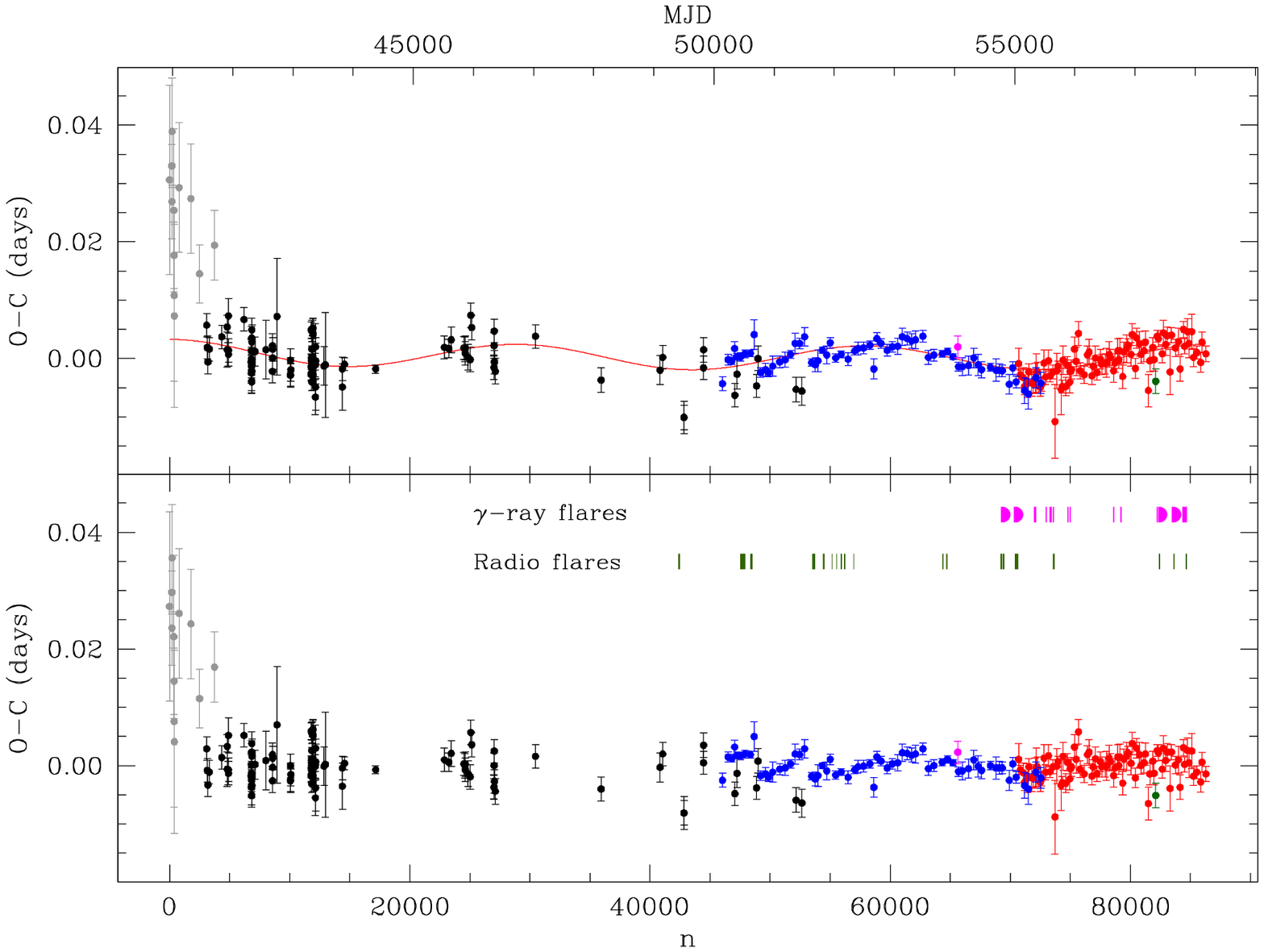}
\caption{Quadratic plus sinusoidal fit to the local epoches. Data points from \cite{leach75} not used in the fit are shown in light gray. Upper plot: residuals of the quadratic term; the solid red line is the best-fit sinusoidal term. Bottom plot: residuals of the full model (quadratic plus sinusoidal terms). The moments of radio ({\em Ryle}/AMI, 15GHz) and $\gamma$-ray ({\em Fermi} LAT, 0.08-300 GeV) flares are also shown. These are taken from \cite{zd16}, \cite{zd18}. }
\label{fig4}
\end{figure}

To fit the sinusoidal variations seen in the residuals of the quadratic model, we used a model consisting of a sum of quadratic and sinusoidal terms:

$$
 T_n = T_0 + P_0n + cn^2 + a_s\sin(\frac{2\pi}{P_s}(n-n_{0s}))
$$

where $c=P_0\dot{P}/2$, $a_s$ is the amplitude of the sine function, $P_s$ id the sine period, $n_{0s}$ is the initial epoche of sinusoidal variations (in orbital cycles). The fit is shown in Fig.~\ref{fig4}, its parameters are listed in Table~\ref{table2}. The $\chi^2$ value of the fit is much smaller that that for the quadratic model, but still too large to formally accept the model. As the remaining deviations of the data from the model look random, to estimate errors of the parameters, we used the same {\em Bootstrapping residuals} method which was used when fitting the template light curve to X-ray data.

At least part of the remaining deviations in Fig.~\ref{fig4} (bottom plot) are clearly due to irregular and abrupt changes in the orbital period. We will discuss these below.

\section{The nature of the sinusoidal component in the period change}\label{sec:sine}

Theoretically speaking, the sinusoidal variations of the orbital period could be explained by three mechanisms:

\begin{enumerate}
 \item Regular changes in the mass loss rate of the WR star.
 \item Apsidal motion in the system.
 \item A presence of a third body.
\end{enumerate}

The first hypothesis is ruled out by simple numerical estimates. Although the wind of the WR star is probably not spherically symmetric (see our subsequent paper), we can roughly relate $\dot{P}$ to $\dot{M}$ by using a relationship for a spherical case: $2\dot{M}/M = \dot{P}/P$, where $M$ is the total mass of a binary and $P$ is the orbital period. Let us assume that $\dot{M}(n)=\dot{M_0}+a_{\dot{M}}\sin(2\pi/P_s(n-n_{0s}))$.  After some simple algebra, the expected amplitude of the (sinusoidal) residuals with the quadratic ephemeris

$$
a_s[{\rm d}]=\frac{P_0[{\rm d}]P_s[{\rm yr}]}{\pi M[{\rm M_\odot}]}a_{{\rm \dot{M}}}[{\rm M_\odot/yr}]
$$

Substituting the values from Table~\ref{table2} and $M=15$~M$_\odot$, we obtain $a_{\rm{\dot{M}}}\simeq 0.03$~M${_\odot}$~yr$^{-1}$. This means that to explain the observed sinusoidal variations of the orbital period, the WR star has to {\em change} its mass loss rate by $0.03$~M$_\odot$~yr$^{-1}$, which is absolutely impossible. More precise estimates taking into account e.g. accretion onto the compact companion can be done following \cite{tout91}, but they would clearly give similar results, as the compact companion can accrete only a small fraction of the WR wind mass. Above all, this hypothesis assumes an underlying mechanism which forces strictly periodic long term changes of the WR mass loss rate. Such mechanisms are currently unknown.

The situation with the second hypothesis is more complicated. First note that apsidal motion as an explanation of the global period change (that is the assumption that the parabolic shape of the residuals with linear ephemeris is in fact a part of sinusoidal variations) is definitely ruled out. The amplitude of such sine function would require the eccentricity of the binary orbit to be much higher than unity.

As for the $15.77$~yr sinusoidal variations, there are two numerical quantities which can be verified in the framework of the apsidal motion model: the eccentricity $e$ and the period of the alledged apsidal motion $U$. Another qualitative feature of the apsidal variability is the shape of the light curve (more precisely, the distance between the primary and secondary minima) which should change within the period $U$.

It is known (see e.g. \citealp{batten73}, page 88) that the amplitude of sinusoidal variations in the apsidal motion model is $a_s\simeq eP/\pi$. Using our values for $a_s$ and $P$, we obtain $e\simeq 0.033$. This value is not extraordinary and cannot be an argument to reject the hypothesis.

The time interval between light curve minima in a system with apsidal motion is given by

$$
 (T_2-T_1)\frac{\pi}{P} - \frac{\pi}{2} = e\cos\omega(1+{\rm cosec}^2i)\, ,
$$

where $\omega$ is the longitude of periastron and $i$ is the orbital inclination. Thus the maximal change of the phase difference between the minima is 

$$
 (T_2-T_1)_{\rm max}/P = 2e(1+{\rm cosec}^2i)/\pi
$$

Substituting $e$ just found above and $i=43\arcdeg$ \cite{zd13}, we obtain \mbox{$(T_2-T_1)_{max}/P \simeq 0.07$}. While there is no secondary minimum in the X-ray light curve of \mbox{Cyg~X-3}, the above value gives a characteristic phase scale for the alledged variations in the light curve. The obtained value is rather small and such variations, if present, could be easily get unnoticed. Thus, the observed stability of the X-ray light curve (except irregular flares) cannot be used as an argument against the hypothesis, either.

Finally, we can verify the compatibility of the observed sine period with the apsidal motion model. \cite{batten73} in Chapter 6 gives the formula (2) for the ratio $P/U$. Making simplifying assumptions that (i)the compact object can be considered as a point source and hence its $k_{22}=0$, (ii)the eccentricity is small so we can neglect all $e^2$ terms, and (iii)the rotation of the WR component is synchronous with the orbital revolution, we come to the formula

$$
 \frac{P}{U} = k_{21}\left(\frac{R_{\rm WR}}{a}\right) ^5\left[16\frac{M_{\rm C}}{M_{\rm WR}}+1\right]\, ,
$$

where $k_{21}$ is the apsidal parameter of the WR component depending on its internal density structure (e.g. the polytrope index). Observations and theory give the interval $k_{21} \simeq 0.001-0.01$ for stars of various types (albeit, non-WR). We will adopt the average value $k_{21}=0.005$ for our rough estimates. Assuming the mass ratio $M_C/M_{\rm WR}$=0.23 \citep{zd13}, the term in the square brackets is equal to 4.68. The primary source of uncertainly clearly comes from the error in the relative radius of the WR component as it enters the formula in the fifth power. To estimate a possible interval of $U$, we computed $U$ for two values of radius of a helium core of the WR star: 1 and 2~R$_\odot$. The distance between the components in \mbox{Cyg~X-3} is about 3.5~R$_\odot$ \citep{vilhu13}. Substituting these values to the equation, we obtain $U_1= 4489$~d ($\sim 12$~yr) and $U_2=140$~d. The obtained values are not surprising as the system is extremely close. This result should be taken with extreme caution given all the uncertainties involved. However, the conclusion is that the apsidal motion hypothesis cannot be ruled out by the available data.

Let us now estimate the parameters of the third body configuration. The distance from the close binary to the center of mass of the alledged triple system is determined by the light equation $a_1\sin i=ca_s$, which results in $a_1=1.08\times 10^{13}$~cm (about $0.72$~au). Making simplifying assumptions that (i)the third body and the close binary orbits lie in the same plane, (ii)the orbit of the third body is circular, and (iii)the mass of the close binary (WR+C) is 15~M$_\odot$, from the third Kepler's law we obtain $M_3\simeq 0.7$~M$_\odot$, the distance of the third body from the center of mass of the triple system is $a_2\simeq 16$~au. These values look reasonable for a hierarchical triple system.

\section{Irregular changes of the residuals}\label{sec:irreg}

The bottom plot in Fig.~\ref{fig4} shows several abrupt residual changes in the interval of MJD 50000-54000. Note that these changes were also present in \cite{singh02}, although they were not as evident as in our study, due to smaller amount of {\em RXTE} ASM data used by the authors.

A tempting explanations is that such changes could be caused by sudden changes in the mass loss rate of the WR star. The mass loss would drop at the change moment and then quickly return to the previous value or actually higher, as the residual curve after the observed drops is steeper than ``normal''. However, a simple imaginary experiment shows that this cannot be the case. First note that decrease of the mass loss rate cannot decrease the period, which will still increase just at a lower pace. Let us suppose that at the moment of a sudden change the mass loss rate dropped to zero. Starting from this moment, the orbital period is constant. This means that the residuals from the linear ephemeris will lie along a straight line. If the estimated period is equal to the exact one, the line will be horizontal. If the estimated period is shorter that the exact one, the slope of the line will be positive, and if it is longer, then the slope will be negative. Thus we must assume that at the moment of a sudden change the estimated period is longer than the exact one. What is the difference between the two? It depends on how much the mass loss rate decreased. Let us for simplicity suppose again that the mass loss rate dropped to zero. Then the difference between the estimated and exact periods is equal to the difference between the residuals before and after the event divided by the number of orbital cycles passed between the two residual points. A typical difference in the residuals at the moments of sudden changes is $0.002$~d, and the number of orbital cycles is $600$. Thus, the difference between the estimated and the exact periods is about $3\times 10^{-6}$~d. If the mass loss rate dropped not to zero but to some larger value, this period difference would be higher. Given the accuracy of the period found in out fits, the difference seems too high (by about 2-3 orders of magnitude). We conclude that sudden changes in the mass loss rate of the WR star is a highly unlikely explanation of the sudden changes in the residuals.

Leaving apart possible calibration issues with {\em RXTE} ASM as unlikely, another possible explanation could be that the shape of the X-ray light curve in this MJD interval had systematic differences with the template curve which resulted in systematic errors in local epoches. Such changes could be caused by some processes which changed accretion onto the compact object. After the end of a ``perturbation'' stage the light curve would return to its ``normal'' shape. On the other hand, visual inspection of the light curves did not show clear signs of such changes. Any unusual processes could possibly manifestate themselves also in \mbox{radio/$\gamma$-ray} outbursts. In Fig.~\ref{fig4} we plot the moments of such outbursts. While the two sequences of events are not exactly coincident, some rough correlation between the two is probably present.

Another highly speculative explanation could be related to asymmetry of mass loss rate through jets. Let us assume that a blob of matter is ejected in one of the jets. The momentum which it carries away would create a kick which would change the orbital angular momentum of the system, resulting in period change. The value $3\times 10^{-6}$~d is then the real period change due to an instant kick. To check this hypothesis, let us make some simplifying assumptions: the orbit is circular and remains circular after the kick, the change in the system mass is negligible. The orbital angular momentum is 

$$
 L=\frac{M_1M_2}{M_1+M_2}\sqrt{G(M_1+M_2)a}
$$

where $M_1$ and $M_2$ are the masses of the components and $a$ is the orbital separation. Thus, 

$$
 \frac{dL}{L}=\frac{1}{2}\frac{da}{a}
$$

A change of the orbital separation is related to a change of the period through Kepler's third law

$$
 \frac{dP}{P}=-\frac{3}{2}\frac{da}{a}
$$

On the other hand,

$$
 dL=a M_{\rm blob}V_{\rm blob}\cos \alpha
$$

where $M_{\rm blob}$ and $V_{\rm blob}$ are the mass and velocity of the blob, and $\alpha$ is the angle between the jet axis and the instant vector of the orbital velocity of the compact object. Combining these equations, we come to the formula for blob mass required to obtain a period change $dP$:

$$
 M_{\rm blob} = -\frac{1}{3}\frac{dP}{P}\frac{L}{aV_{\rm blob}\cos\alpha}
$$

Assigning reasonable values \mbox{$M_1=12$~M$_\odot$},  \mbox{$M_2=3$~M$_\odot$}, \mbox{$a=3.5$~R$_\odot$}, \mbox{$\alpha=37.3\arcdeg$} (from \citealp{zd18}, $dP=3\times 10^{-6}$~d, $P=0.2$~d, and taking $V_{\rm blob}$ to be equal to the jet velocity ($0.5c$, \citealp{marti01}), we obtain $M_{\rm blob} \simeq 0.9\times 10^{-7}$~M$_\odot$. The mass loss rate of the WR star is about $10^{-5}$~M$_\odot$~yr$^{-1}$, the accretion rate is probably $\leq 0.01$ of this value. Thus, it seems practically impossible that a blob would have mass accreted by the compact source during a year. 

\section{Long-term variability of X-ray flux}\label{sec:longterm}

\begin{figure}
\centering
\plotone{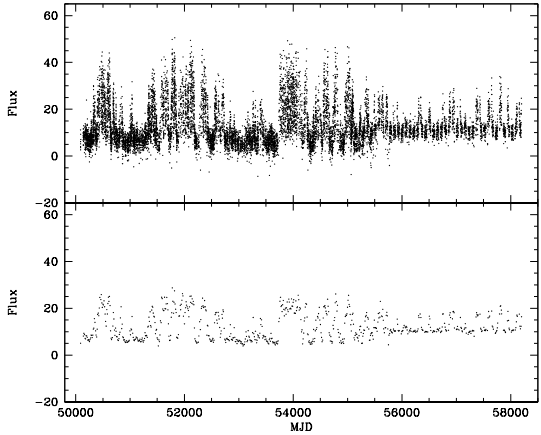}
\caption{A combined {\em RXTE} ASM and {\em MAXI} light curve of \mbox{Cyg~X-3}.Top panel: original data. Bottom panel: average light curve (10 days bins), including only positive fluxes.}
\label{fig5}
\end{figure}

The continuous monitoring of \mbox{Cyg~X-3} by {\em RXTE} ASM and {\em MAXI} provides an opportunity to search for a long term periodic variability of X-ray flux. Such search is motivated by the fact that the system has jets \citep{zd18} and hence, possibly a small accretion disk. Precession of the jets/disk should lead to precessional flux variations. Indeed, \cite{zd18} estimated the period of jet precession \mbox{($\sim 170$~d)} from the \mbox{20 s} difference between the orbital period and the period of radio emission, which they interpreted as a beat of the orbital modulation with jet precession.

To combine the {\em RXTE} ASM and {\em MAXI} data into a single sequence, we scaled the latters so that their mean and standard deviation are equal to those of {\em RXTE} ASM data. A plot of the resulting light curve as a function of Julian date is shown in Fig.~\ref{fig5}, top panel. It is immediately clear that strong irregular flux changes would probably mask any regular variability with amplitude presumably much smaller than the irregular variations. Also, some {\em RXTE} ASM fluxes are negative, probably due to errors in background subtraction during processing. Thus, we selected only positive fluxes and computed an average light curve using 10 days bins. The resulting light curve is shown in the bottom panel.

The average light curve seems to indicate two states of the object: a low state when the flux is nearly constant and a high state where the flux is strongly variable. As we shall see below, this is not just an impression. Also, a linear trend in the data over the period of 18 years is clearly seen. Thus, when computing Fourier spectra, we subtracted the trend. To check for consistency of Fourier analysis, we repeated all calculations described below, for 20 days binning, and obtained the same results.

\begin{figure}
\centering
\plotone{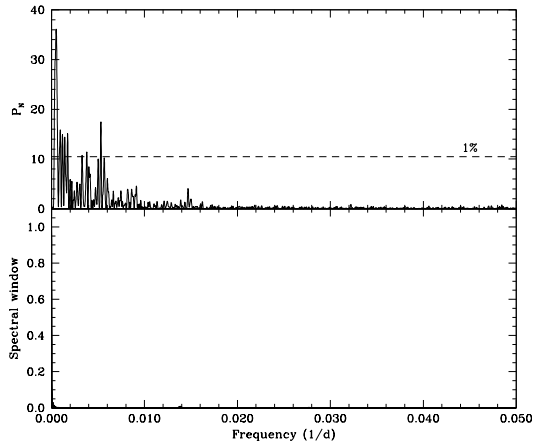}
\caption{Top panel: Fourier spectrum of {\em RXTE} ASM and {\em MAXI} X-ray flux. A horizontal dashed line shows a 1 per cent significance level. Bottom panel: spectral window.}
\label{fig6}
\end{figure}

A discrete Fourier spectrum obtained with the average light curve is shown in Fig.~\ref{fig6}. The spectrum is computed up to the Nyiquist frequency, which is approximately equal to $0.05$ for 10 days binning. The spectral window is featureless so ``cleaning'' the discrete Fourier spectrum was not performed.

\begin{figure}
\centering
\plotone{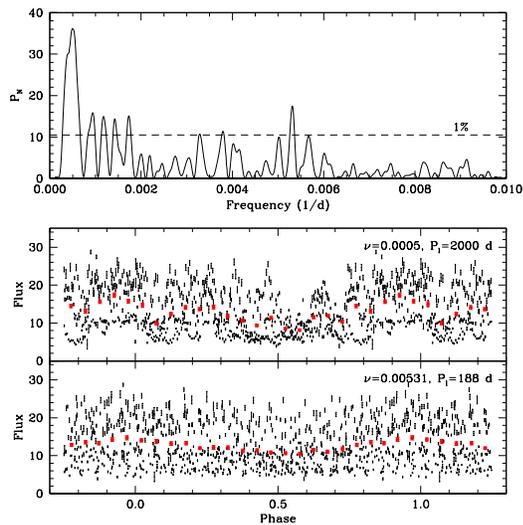}
\caption{Top panel: Fourier spectrum in the $0-0.01$ frequency interval. Two bottom panels: convolved light curves corresponding to the two most prominent peaks in the Fourier spectrum. Individual measurements are shown by small black dots (errors are not shown to avoid clutter), the mean convolved light curve with phase bins 0.05 is shown by large red dots.}
\label{fig7}
\end{figure}

In Fig.~\ref{fig7}, a part of the Fourier spectrum for the frequency interval $0-0.01$ is shown, along with two convolved light curves corresponding to the two most prominent peaks in the spectrum. The period of the highest peak is $P_l=2000$~d. While being formally significant, the correspoding convolved light curve is rather noisy. Also, the period roughly equals to the distance between the the most prominent activity intervals in Fig.~\ref{fig5}. On the other hand, the second highest peak ($P_l=188$~d) produces a much smoother convolved light curve. A striking feature of the convolved light curve (already seen in Fig.~\ref{fig5}) is that it consists of two components above and below the flux value $\sim 14$~counts~s$^{-1}$, with a clear gap between the two. It gives an impression that long-term periodic variability only exists in the high state of \mbox{Cyg~X-3}. To check this assumption, we computed two Fourier spectra for fluxes above and below the threshold equal to 14. 

The first spectrum is very similar to the one shown in Fig.~\ref{fig7}. The four peaks in the $0.0007-0.0018$ frequency intervals have greatly decreased and became insignificant, while the two highest peaks remained in the same positions. The second spectrum shows no prominent peaks, being noisy and located well below the 1 per cent significance level.

This behavior can have two explanations. The first explanation is that in the low state, the fluxes are too small and their errors are too large to allow for detection of any variability. The second explanation is physical. In the low state, the accretion rate to the compact companion may be small so no accretion disk is formed (Bondi-Hoyle accretion). In the high state, a small precessing accretion disk may be formed.

\section{Conclusions}\label{sec:concl}

The available literature on the period change of \mbox{Cyg~X-3} provides 115 epoches of X-ray minima. By using archival observations of {\em RXTE} ASM, {\em MAXI}, {\em SUZAKU} and {\em AstroSat}, we added 155 new epoches to the list. Our analysis of these data yields the most accurate estimate of the rate of period change to date. 

We do not confirm existence of a second derivative of the orbital period suggested by some authors earlier. Instead, we find that the changes in the period can be fit by a sum of quadratic and sinusoidal functions. The period of sinusoidal variations is $15.8$~yr. They can be related either to apsidal motion in the close binary with eccentricity $e\simeq 0.03$ or by a presence of a third body with the mass of about $0.7$~M$_\odot$ located at a distance $\sim 16$~au from the close binary.

We also detect irregular and abrupt changes in the residuals between the best fit ephemeris and the data. Their origin remains unclear.

Large amount of data and nearly continuous time coverage of {\em RXTE} ASM and {\em MAXI} light curves allowed us to search for long term variability of Cyg X-3. A tentative period of about 188 days was found. The period is present only in high state of the binary. Such a period could be attributed to a small precessing disk around the compact object.

\acknowledgments

We thank Prof. K.A. Postnov for valuable discussions. This research has made use of {\em RXTE} ASM and {\em SUZAKU} data obtained through the High Energy Astrophysics Science Archive Research Center Online Service, provided by the NASA/Goddard Space Flight Center, and also of the data from the {\em AstroSat} mission of the Indian Space Research Organization (ISRO), archived at the Indian Space Science Data Centre (ISSDC). This research has also made use of the {\em MAXI} data provided by RIKEN, JAXA and the {\em MAXI} team. Authors acknowledge support by the Russian Science Foundation grant 17-12-01241.

\end{document}